# Energy- and flux-budget (EFB) turbulence closure model for the stably stratified flows. Part I: Steady-state, homogeneous regimes

S. S. Zilitinkevich[1,2,3], T. Elperin[4], N. Kleeorin[4] and I. Rogachevskii[4]

[1] Division of Atmospheric Sciences, University of Helsinki, Finland

[2] Finnish Meteorological Institute, Helsinki, Finland

[3] Nansen Environmental and Remote Sensing Centre / Bjerknes Centre for Climate Research, Bergen, Norway

[4] Pearlstone Center for Aeronautical Engineering Studies, Department of Mechanical Engineering, Ben-Gurion University of the Negev, Beer-Sheva, Israel

# Energy- and flux-budget (EFB) turbulence closure model for the stably stratified flows. Part I: Steady-state, homogeneous regimes

S. S. Zilitinkevich[1,2,3], T. Elperin[4], N. Kleeorin[4] and I. Rogachevskii[4]

[1]Division of Atmospheric Sciences, University of Helsinki, Finland

[2]Finnish Meteorological Institute, Helsinki, Finland

[3]Nansen Environmental and Remote Sensing Centre / Bjerknes Centre for Climate Research, Bergen, Norway

[4]Pearlstone Centre for Aeronautical Engineering Studies, Department of Mechanical Engineering, Ben-Gurion University of the Negev, Beer-Sheva, Israel

## Abstract

We propose a new turbulence closure model based on the budget equations for the key second moments: turbulent kinetic and potential energies: TKE and TPE (comprising the turbulent total energy: TTE = TKE + TPE) and vertical turbulent fluxes of momentum and buoyancy (proportional to potential temperature). Besides the concept of TTE, we take into account the non-gradient correction to the traditional buoyancy flux formulation. The proposed model grants the existence of turbulence at any gradient Richardson number, Ri. Instead of its critical value separating – as usually assumed – the turbulent and the laminar regimes, it reveals a transition interval, 0.1< Ri <1, which separates two regimes of essentially different nature but both turbulent: strong turbulence at Ri<<1; and weak turbulence, capable of transporting momentum but much less efficient in transporting heat, at Ri>1. Predictions from this model are consistent with available data from atmospheric and lab experiments, direct numerical simulation (DNS) and large-eddy simulation (LES).

**Keywords:**  Turbulence closure   Stable stratification   Critical Richardson number   Kinetic, potential and total turbulent energies   Turbulent fluxes   Eddy viscosity and heat-conductivity   Anisotropy of turbulence   Turbulent length scale   Turbulent boundary layers

# 1. Introduction

Most of the practically used turbulence closure models are based on the concept of the down-gradient transport. Accordingly they express turbulent fluxes of momentum and scalars as products of the mean gradient of the transported property and the corresponding turbulent transport coefficient (eddy viscosity, $K_M$, heat conductivity, $K_H$, or diffusivity, $K_D$). Following Kolmogorov (1941), the latter are taken proportional to the turbulent velocity scale, $u_T$, and length scale, $l_T$:

$$K_M \sim K_H \sim K_D \sim u_T l_T. \qquad (1)$$

Usually $u_T^2$ is identified with the turbulent kinetic energy (TKE) per unit mass, $E_K$. The latter is calculated from the TKE budget equation using the Kolmogorov's closure for the TKE dissipation rate:

$$\varepsilon_K \sim E_K / t_T, \qquad (2)$$

where $t_T \sim l_T / u_T$ is the turbulent dissipation time scale.

This approach is justified when applied to the neutral-stability flows (where $l_T$ can be taken proportional to the distance form the nearest wall).

However, it faces difficulties in stratified flows (both stable and unstable). The turbulent Prandtl number $\mathrm{Pr}_T = K_M / K_H$ exhibits essential dependence on the stratification and cannot be considered as constant. Next, as follows from the budget equations for the vertical turbulent fluxes, the velocity scale $u_T$ characterising vertical turbulent transports is determined as the root mean square (r.m.s.) vertical velocity $u_T \sim \sqrt{E_z}$ (where $E_z$ is the energy of the vertical velocity fluctuations). In neutral stratification $E_z \sim E_K$, which is why the traditional formula $u_T \sim \sqrt{E_K}$ holds true. But in strongly stable stratification it becomes insufficiently accurate because of stability dependence of the anisotropy of turbulence $A_z \equiv E_z / E_K$: $A_z$ generally decreases with increasing stability.

To reflect the effect of stratification, the turbulent length scales for the momentum, $l_{TM}$, and heat, $l_{TH}$, are taken different. As a result the above closure scheme (formulated by Kolmogorov for the neutral stratification and well-grounded in this and only in this case) loses its constructiveness: the unsolved part of the problem is simply displaced from $\{K_M, K_H\}$ to $\{l_{TM}, l_{TH}\}$. The TKE budget equation becomes insufficient to determine additional unknown parameters.

Numerous alternative turbulence closures were formulated using equations for other turbulent parameters (besides TKE) together with heuristic hypotheses and empirical relationships, but no consensus is still achieved; see overviews by Weng and Taylor (2003), and Umlauf and Burchard (2005).

In this paper we analyse the effects of the density stratification on the turbulent energies and vertical turbulent fluxes in the stably stratified atmospheric (or oceanic) boundary-layer flows, in which the horizontal variations of the mean velocity and temperature are much weaker than the vertical variations. The proposed theory provides realistic stability dependencies of the turbulent Prandtl number, the vertical anisotropy, and the vertical turbulent length scale. Below we present material in meteorological terms, but all results can be easily reformulated in terms of water currents in the ocean or lakes, expressing the buoyancy through the temperature and salinity instead of the temperature and humidity.

We consider a minimal set of the budget equations for the second order moments, namely those for the vertical fluxes of buoyancy (proportional to potential temperature) and momentum, the TKE and the turbulent potential energy, TPE (proportional to the mean squared potential temperature fluctuation). In these equations we account for some usually neglected features but leave more detailed treatment of the third-order transports and the pressure-velocity correlations for future analysis. In particular, we advance the familiar "return to isotropy" model to more realistically determine the stability dependence of the vertical anisotropy, $A_z$. We also take into account a non-gradient correction to the traditional, down-gradient formulation for the turbulent flux of potential temperature. This approach allows deriving a reasonably simple turbulence closure scheme including realistic energy budgets and stability dependence of $\Pr_T$.

We consider the total (kinetic + potential) turbulent energy (TTE), derive the TTE budget equation, and demonstrate that the TTE in stably stratified sheared flows does not completely decay even in very strong static stability. This conclusion, deduced from the general equations independently of the concrete formulation for the turbulent length scale, argues against the widely recognised concept of the critical Richardson number.

Recall that the Richardson number, Ri, is defined as the squared ratio of the Brunt-Väisälä frequency, $N$, to the velocity shear, $S$:

$$\text{Ri} = \left(\frac{N}{S}\right)^2, \quad S^2 = \left(\frac{\partial U}{\partial z}\right)^2 + \left(\frac{\partial V}{\partial z}\right)^2, \quad N^2 = \beta \frac{\partial \Theta}{\partial z}, \qquad (3)$$

where $z$ is the vertical co-ordinate, $U$ and $V$ are the mean velocity components along the horizontal axes $x$ and $y$, $\Theta$ is the mean potential temperature, $\beta = g/T_0$ is the buoyancy parameter, $g$ =9.81 m s$^{-1}$ is the acceleration due to gravity, and $T_0$ is a reference value of the absolute temperature. As proposed by Richardson (1920), it quantifies the effect of static stability on turbulence. Since that time, a principal question "whether or not the stationary turbulence can be maintained by the velocity shear at very large Richardson numbers" has been the focus of attention in the theory of stably stratified turbulent flows.

A widely recognised opinion is that turbulence decays when Ri exceeds some critical value, $Ri_c$ (with the frequently quoted estimate of $Ri_c = 0.25$). However, the concept of critical Ri was neither rigorously derived from basic physical principles nor demonstrated empirically. Just opposite, it contradicts to long standing experimental evidence (see detailed discussion in Zilitinkevich et al., 2007).

It is worth emphasising that turbulence closure models based on the straightforward application of the TKE budget equation and Kolmogorov's closure hypotheses, Eqs. (1) and (2), do imply the existence of $Ri_c$. In practical atmospheric modelling such closures are not acceptable. In particular, they lead to unrealistic decoupling of the atmosphere from the underlying surface in each case when Ri in the surface layer exceeds $Ri_c$. Since the milestone paper of Mellor and Yamada (1974), to prevent undesirable appearance of $Ri_c$, turbulence closures in practical use are equipped with correction-coefficients specifying the ratios $K_M (u_T l_T)^{-1}$ and $K_H (u_T l_T)^{-1}$ as essentially different single-valued functions of Ri. Usually these functions are not derived in the context of the closure in use, but either determined empirically or taken from independent theories. Using this trick, modellers close eyes on the fact that corrections could be inconsistent the basic closure model formalism.

## 2. Reynolds equations and budget equations for second moments

We consider atmospheric flows, in which typical variations of the mean wind velocity $\mathbf{U} = (U_1, U_2, U_3) = (U, V, W)$ and potential temperature $\Theta$ (or virtual potential temperature involving specific humidity) in the vertical [along $x_3$ (or $z$) axis] are much larger than in the horizontal [along $x_1, x_2$ (or $x,y$) axis], so that the terms proportional to their horizontal gradients in the budget equations for turbulent statistics can be neglected. $\Theta$ is defined as $\Theta = T(P_0/P)^{1-1/\gamma}$, where $T$ is the absolute temperature, $P$ is the pressure, $P_0$ is its reference value, and $\gamma = c_p / c_v = 1.41$ is the specific heats ratio.

We also assume that the vertical scale of motions (limited to the height scale of the atmosphere or the ocean: $H \sim 10^4$ m) is much smaller than their horizontal scale, so that the mean flow vertical velocity is typically much smaller than the horizontal velocity. In this context, to close the Reynolds equations we need only the vertical component, $F_z$, of the potential temperature flux, $F_i$, and the two components of the Reynolds stresses, $\tau_{ij}$, namely those representing the vertical turbulent flux of momentum: $\tau_{13}$ and $\tau_{23}$.

The mean flow is described by the momentum equations:

$$\frac{DU_1}{Dt} = f U_2 - \frac{1}{\rho_0} \frac{\partial P}{\partial x} - \frac{\partial \tau_{13}}{\partial z}, \tag{4}$$

$$\frac{DU_2}{Dt} = -fU_1 - \frac{1}{\rho_0}\frac{\partial P}{\partial y} - \frac{\partial \tau_{23}}{\partial z}, \quad (5)$$

and the thermodynamic energy equation:

$$\frac{D\Theta}{Dt} = -\frac{\partial F_z}{\partial z} + J, \quad (6)$$

where $D/Dt = \partial/\partial t + U_k \partial/\partial x_k$; $\tau_{ij} = \langle u_i u_j \rangle$; $F_i = \langle u_i \theta \rangle$; $t$ is the time; $f = 2\Omega \sin\varphi$, $\Omega_i$ is the earth's rotation vector parallel to the polar axis ($|\Omega_i| \equiv \Omega = 0.76 \cdot 10^{-4}$ s$^{-1}$); $\varphi$ is the latitude, $\rho_0$ is the mean density; $J$ is the heating/cooling rate ($J=0$ in adiabatic processes); $P$ is the mean pressure; $\mathbf{u} = (u_1, u_2, u_3) = (u, v, w)$ and $\theta$ are the velocity and potential-temperature fluctuations; the angle brackets denote the ensemble average [see Holton (2004) or Kraus and Businger (1994)].

The budget equations for the TKE, $E_K = \frac{1}{2}\langle u_i u_i \rangle$, the "energy" of the potential temperature fluctuations, $E_\theta = \frac{1}{2}\langle \theta^2 \rangle$, the potential-temperature flux, $F_i = \langle u_i \theta \rangle$ [with the vertical component $F_3 = F_z = \langle w\theta \rangle$], and the Reynolds stress $\tau_{ij} = \langle u_i u_j \rangle$ [with the components $\tau_{i3} = \langle u_i w \rangle$ ($i$=1,2) representing the vertical flux of momentum] can be found, e.g., in Kaimal and Fennigan (1994), Kurbatsky (2000) and Cheng et al. (2002):

$$\frac{DE_K}{Dt} + \nabla \cdot \mathbf{\Phi}_K = -\tau_{ij}\frac{\partial U_i}{\partial x_j} + \beta F_z - \varepsilon_K \quad (7a)$$

or approximately

$$\frac{DE_K}{Dt} + \frac{\partial \Phi_K}{\partial z} \approx -\tau_{i3}\frac{\partial U_i}{\partial z} + \beta F_z - \varepsilon_K, \quad (7b)$$

$$\frac{DE_\theta}{Dt} + \nabla \cdot \mathbf{\Phi}_\theta = -F_z\frac{\partial \Theta}{\partial z} - \varepsilon_\theta \quad (8a)$$

or approximately

$$\frac{DE_\theta}{Dt} + \frac{\partial \Phi_\theta}{\partial z} = -F_z\frac{\partial \Theta}{\partial z} - \varepsilon_\theta, \quad (8b)$$

$$\frac{DF_i}{Dt} + \frac{\partial}{\partial x_j}\Phi_{ij}^{(F)} = \beta_i \langle \theta^2 \rangle + \frac{1}{\rho_0}\langle \theta \nabla_i p \rangle - \tau_{ij}\frac{\partial \Theta}{\partial z}\delta_{j3} - F_j\frac{\partial U_i}{\partial x_j} - \varepsilon_i^{(F)}, \quad (9a)$$

and for $F_3 = F_z$

$$\frac{DF_z}{Dt} + \frac{\partial}{\partial z}\Phi_F = \beta\langle\theta^2\rangle + \frac{1}{\rho_0}\langle\theta\frac{\partial}{\partial z}p\rangle - \langle w^2\rangle\frac{\partial\Theta}{\partial z} - \varepsilon_z^{(F)}$$

$$\approx C_\theta\beta\langle\theta^2\rangle - \langle w^2\rangle\frac{\partial\Theta}{\partial z} - \varepsilon_z^{(F)}, \tag{9b}$$

$$\frac{D\tau_{ij}}{Dt} + \frac{\partial}{\partial x_k}\Phi_{ijk}^{(\tau)} = -\tau_{ik}\frac{\partial U_j}{\partial x_k} - \tau_{jk}\frac{\partial U_i}{\partial x_k} + \left[\beta(F_j\delta_{i3} + F_i\delta_{j3}) + Q_{ij} - \varepsilon_{ij}^{(\tau)}\right] \tag{10a}$$

and for $\tau_{i3}$ ($i=1,2$)

$$\frac{D\tau_{i3}}{Dt} + \frac{\partial}{\partial z}\Phi_i^{(\tau)} = -\langle w^2\rangle\frac{\partial U_i}{\partial z} - \left[-\beta F_i - Q_{i3} + \varepsilon_{i3}^{(\tau)}\right] \approx -\langle w^2\rangle\frac{\partial U_i}{\partial z} - \varepsilon_{i3}, \tag{10b}$$

where $\beta_i = \beta e_i$ and $\mathbf{e}$ is the vertical unit vector, $F_i = \langle u_i\theta\rangle$ ($i=1,2$) are the horizontal fluxes of potential temperature, $-\tau_{ij}\partial U_i/\partial x_j$ is the TKE production rate; $\delta_{ij}$ is the unit tensor ($\delta_{ij}=1$ for $i=j$ and $\delta_{ij}=0$ for $i\neq j$).

$\Phi_K$, $\Phi_\theta$ and $\Phi_F$ are the third-order moments representing the turbulent transports of the TKE and the "energy" of potential temperature fluctuations:

$$\mathbf{\Phi}_K = \frac{1}{\rho_0}\langle p\,\mathbf{u}\rangle + \frac{1}{2}\langle u^2\,\mathbf{u}\rangle \quad \text{that is} \quad \Phi_K = \frac{1}{\rho_0}\langle p\,w\rangle + \frac{1}{2}\langle u^2\,w\rangle, \tag{11a}$$

$$\mathbf{\Phi}_\theta = \frac{1}{2}\langle\theta^2\,\mathbf{u}\rangle, \quad \text{that is} \quad \Phi_\theta = \frac{1}{2}\langle\theta^2\,w\rangle, \tag{11b}$$

and the turbulent transports of the fluxes of potential temperature and momentum (the fluxes of fluxes):

$$\Phi_{ij}^{(F)} = \frac{1}{2\rho_0}\langle p\,\theta\rangle\delta_{ij} + \langle u_i u_j\theta\rangle,\quad \Phi_{33}^{(F)} = \Phi_F = \frac{1}{2\rho_0}\langle p\,\theta\rangle + \langle w^2\,\theta\rangle, \tag{12}$$

$$\Phi_{ijk}^{(\tau)} = \langle u_i u_j u_k\rangle + \frac{1}{\rho_0}\left(\langle pu_i\rangle\delta_{jk} + \langle pu_j\rangle\delta_{ik}\right), \tag{13a}$$

$$\Phi_{i33}^{(\tau)} = \Phi_i^{(\tau)} = \langle u_i w^2\rangle + \frac{1}{\rho_0}\langle pu_i\rangle, \quad (i=1,2). \tag{13b}$$

$Q_{ij}$ are correlations between the fluctuations of pressure, $p$, and the velocity shears:

$$Q_{ij} = \frac{1}{\rho_0}\left\langle p\left(\frac{\partial u_i}{\partial x_j} + \frac{\partial u_j}{\partial x_i}\right)\right\rangle. \tag{14}$$

$\varepsilon_K$, $\varepsilon_{ij}^{(\tau)}$, $\varepsilon_\theta$ and $\varepsilon_i^{(F)}$ are operators including molecular constants:

$$\varepsilon_K = \nu\left\langle\frac{\partial u_i}{\partial x_k}\frac{\partial u_i}{\partial x_k}\right\rangle, \quad \varepsilon_{ij}^{(\tau)} = 2\nu\left\langle\frac{\partial u_i}{\partial x_k}\frac{\partial u_j}{\partial x_k}\right\rangle, \tag{15a}$$

$$\varepsilon_\theta = -\kappa\langle\theta\,\Delta\theta\rangle, \quad \varepsilon_i^{(F)} = -\kappa\left(\langle u_i\,\Delta\theta\rangle + \Pr\langle\theta\,\Delta u_i\rangle\right), \tag{15b}$$

where $\nu$ is the kinematic viscosity, $\kappa$ is the temperature conductivity, and $\Pr = \nu/\kappa$ is the Prandtl number. Of these terms essentially positive ones: $\varepsilon_K$, $\varepsilon_{ii}^{(\tau)}$ (that is the diagonal elements $\varepsilon_{11}^{(\tau)}, \varepsilon_{22}^{(\tau)}, \varepsilon_{33}^{(\tau)}$), $\varepsilon_\theta$ and $\varepsilon_i^{(F)}$ represent the dissipation rates for $E_K$, $\tau_{ii}$, $E_\theta$ and $F_i^{(F)}$, respectively. Following Kolmogorov (1941), they are taken proportional to the ratios of the dissipating statistical moment to the turbulent dissipation time scale, $t_T$:

$$\varepsilon_K = \frac{E_K}{C_K\,t_T}, \quad \varepsilon_{ii}^{(\tau)} = \frac{\tau_{ii}}{C_K\,t_T}, \quad \varepsilon_\theta = \frac{E_\theta}{C_P\,t_T}, \quad \varepsilon_i^{(F)} = \frac{F_i}{C_F\,t_T}, \tag{16}$$

where $C_K$, $C_P$ and $C_F$ are dimensionless constants.

Physical mechanisms of dissipation of the non-diagonal components of the Reynolds stress, $\tau_{ij}$ ($i \neq j$), are more complicated. The terms $\varepsilon_{ij}^{(\tau)} = 2\nu\left\langle\frac{\partial u_i}{\partial x_k}\frac{\partial u_j}{\partial x_k}\right\rangle$ in Eq. (10b) are comparatively small and even not necessarily positive, whereas the dissipative role is to a large extent performed by the pressure – shear correlations and the horizontal turbulent transport of the potential temperature. Moreover our analysis does not account for the vertical transport of momentum (that is for the contribution to $\tau_{i3}$) due to internal gravity waves [see, e.g., Section 9.4 in Holton (2004)]. Leaving detailed analyses of the $\tau_{i3}$ budget for future work, we introduce for the Reynolds stress an "effective dissipation rate":

$$\varepsilon_{i3(\text{eff})} \equiv \varepsilon_{i3}^{(\tau)} - \beta\,F_i - Q_{i3} + \text{(unaccounted factors)}, \quad i=1,2; \tag{17}$$

and apply to it the Kolmogorov's closure hypothesis: $\varepsilon_{i3(\text{eff})} \sim \tau_{i3}/t_\tau$, where $t_\tau$ is an "effective dissipation time scale" [the term $\varepsilon_{i3}^{(\tau)}$ is estimated as $\varepsilon_{i3}^{(\tau)} \sim O(\text{Re}^{-1/2})$ and can be neglected]. Accounting for the difference between $t_\tau$ and Kolmogorov's dissipation time scale, $t_T$ [see Eq. (16)], our effective dissipation rates become

$$\varepsilon_{i3(\text{eff})} = \frac{\tau_{i3}}{\Psi_\tau t_T}, \tag{18}$$

where $\Psi_\tau = t_\tau / t_T$ is an empirical dimensionless coefficient. There are no grounds to take it constant: $\Psi_\tau$ could depend on the static stability but is neither zero nor infinite. It is also conceivable that its stability dependence should be monotonic.

In further analysis we employ the approximate version of Eq. (9b). As shown in Appendix A, the second term on the r.h.s. of Eq. (9b), namely $\rho_0^{-1}\langle\theta\,\partial p/\partial z\rangle$, is essentially negative and scales as $\beta\langle\theta^2\rangle$. On these grounds, in its approximate version the sum $\beta\langle\theta^2\rangle + \rho_0^{-1}\langle\theta\,\partial p/\partial z\rangle$ is substituted by $C_\theta\beta\langle\theta^2\rangle$, where $C_\theta < 1$ is an empirical dimensionless constant.

## 3. Turbulent energies

Consider, first, the concept of turbulent potential energy (TPE). Using the state equation and the hydrostatic equation, the density and the buoyancy in the atmosphere are expressed through the potential temperature, $\theta$, and specific humidity, $q$ (in the ocean, through $\theta$ and salinity, $s$). These variables are adiabatic invariants conserved in the vertically displaced portions of fluid, so that the density is also conserved. This allows calculating its fluctuation: $\rho' = (\partial\rho/\partial z)\delta z$ and the fluctuation of potential energy per unit mass:

$$\delta E_P = \frac{g}{\rho_0} \int_z^{z+\delta z} \rho'\,\mathrm{d}z = \frac{1}{2}\frac{b'^2}{N^2}. \tag{19}$$

For certainty, we consider the thermally stratified atmosphere, where the buoyancy, $b$, is expressed through the potential temperature: $b = \beta\theta$. It follows that the turbulent potential energy (TPE) is proportional to the energy of the potential temperature fluctuations:

$$E_p = \left(\frac{\beta}{N}\right)^2 E_\theta = \frac{1}{2}\left(\frac{\beta}{N}\right)^2 \langle\theta^2\rangle. \tag{20}$$

Then multiplying Eq. (8b) by $(\beta/N)^2 = (\partial\Theta/\partial z)^{-1}$ and assuming that $N$ changes only slowly compared to turbulent variations, gives the following TPE budget equation[1]:

$$\frac{DE_P}{Dt} + \frac{\partial}{\partial z}\Phi_P = -\beta F_z - \varepsilon_P = -\beta F_z - \frac{E_P}{C_P t_T}, \qquad (21)$$

where $\Phi_P = (\beta/N)^2 \Phi_\theta$ and $\varepsilon_P = (\beta/N)^2 \varepsilon_\theta$. The term $\beta F_z$ appears in Eqs. (7b) and (21) with opposite signs and describes the energy exchange between TKE and TPE.

The sum of the TKE and TPE is nothing but the total turbulent energy (TTE):

$$E = E_K + E_P = \frac{1}{2}\left(\langle \mathbf{u}^2 \rangle + \left(\frac{\beta}{N}\right)^2 \langle \theta^2 \rangle \right). \qquad (22)$$

Its budget equation is immediately derived summing up Eqs. (7b) and (21). Generally speaking, the time-scale constants $C_K$ and $C_P$ in Eq. (16) characterising the kinetic- and the potential-energy dissipation rates can differ (see Zilitinkevich et al., 2007). We leave analysis of their possible difference for future work and take for simplicity $C_K = C_P$. Then the TTE budget equation becomes

$$\frac{DE}{Dt} + \frac{\partial}{\partial z}\Phi_T = -\tau_{i3}\frac{\partial U_i}{\partial z} - \frac{E}{C_K t_T}, \qquad (23)$$

where $\Phi_T = \Phi_K + \Phi_P$ is the TTE flux divergence.

In the steady state, Eq. (23) reduces to a simple balance between the TTE production $= \tau S$ (where $\tau^2 = \tau_{13}^2 + \tau_{23}^2$) and the TTE dissipation $\sim E t_T^{-1}$, which yields $E \sim \tau S t_T$. In Section 5 we demonstrate that at very large Ri the ratios $\tau/E$, $E_K/E$ and $E_z/E_K$ tend to non-zero constantans. Then estimating $t_T$ through the turbulent length scale, $l_z$, as $t_T \sim l_z E_z^{-1/2} \sim l_z E^{-1/2}$ yields the asymptotic large-Ri estimate: $E \sim (l_z S)^2 > 0$. This reasoning leaves no room for the critical Richardson number.

Traditional analyses of the turbulent energy were basically limited to the TKE budget, Eq. (7b); whereas Eq. (8b) for the squared potential temperature fluctuations, although well-known over decades, was ignored in the operationally used turbulent closure models. Only rather recently $E_\theta$ was treated in terms of the turbulent potential energy (TPE) by Dalaudier and Sidi (1987), Hunt et al. (1988), Canuto and Minotti (1993), Schumann and Gerz (1995), Hanazaki and Hunt (1996, 2004), Keller and van Atta (2000), Stretch et al. (2001), Canuto et al. (2001), Cheng et al. (2002), Luyten et al. (2002, p. 257), Jin et al. (2003), Umlauf, L., (2005) and Rehmann and Hwang (2005). Zilitinkevich (2002) employed the TKE and the TPE budget equations on equal terms to derive an energetically consistent turbulent closure model avoiding the traditional hypothesis $K_H \sim K_M \sim E_K t_T$ (that leads to the dead end, at least in the stable stratification). All three budgets, for TKE, TPE and TTE were considered by Canuto and Minotti (1993) and Elperin et al. (2002).

---

[1] Alternatively the TPE budget equation can be derived from the equation for the fluctuation of buoyancy, $b$, namely, multiplying this equation by $bN^{-2}$, and then applying statistical averaging. It follows that Eq. (21) holds true independently of the assumption that $N$ changes slowly.

## 4. Local model for the steady-state, homogeneous regime

### 4.1. ANISOTROPY OF TURBULENCE

In this section we consider the equilibrium turbulence regime and neglect the third-order transport terms, so that the left hand sides (l.h.s.) in all budget equations become zero; and limit our analysis to the boundary-layer type flows, in which the horizontal gradients of the mean velocity and temperature are negligibly small. These are just the conditions, at which the TKE production rate becomes

$$\Pi = -\boldsymbol{\tau} \cdot \frac{\partial \mathbf{U}}{\partial z} = \tau S, \qquad (24)$$

where $\boldsymbol{\tau} = (\tau_{xz}, \tau_{yz}, 0)$, and $\tau \equiv |\boldsymbol{\tau}|$. It goes without saying that $\Pi$ is determined differently in other types of turbulent flows, in particular in the wave boundary layer below the ocean surface or in the capping inversion layer above the long-lived atmospheric stable boundary layer, where the TKE is at least partially produced by breaking of the surface waves in water or internal gravity waves in the atmosphere (these mechanisms are more close to the oscillating-grid generation of turbulence rather than to its shear generation).

Taking $C_P = C_K$ [see discussion of Eq. (23) in Section 3], Eqs. (19)-(23) yield the following expressions for the turbulent energies:

$$\begin{aligned} E &= C_K t_T \Pi, \quad E_P = -C_K t_T \beta F_z = E \operatorname{Ri}_f, \\ E_K &= C_K t_T (\Pi + \beta F_z) \equiv C_K t_T \Pi (1 - \operatorname{Ri}_f) = E(1 - \operatorname{Ri}_f), \end{aligned} \qquad (25)$$

where $\operatorname{Ri}_f$ is the familiar flux Richardson number defined as the ratio of the TKE consumption for overtaking the buoyancy forces to its production by the velocity shear:

$$\operatorname{Ri}_f \equiv -\frac{\beta F_z}{\Pi} = \frac{\operatorname{Ri}}{\operatorname{Pr}_T} = \frac{E_P}{E}. \qquad (26)$$

As follows from the above analysis, $Ri_f$ is nothing but the ratio of the TPE to the TTE. This interesting fact was overlooked until present[2]. $Ri_f$ is equal to zero in neutral stratification, monotonically increases with increasing stability, but obviously cannot exceed unity. Hence in the very strong static stability (at $Ri \to \infty$) it must approach a non-zero, positive limit, $Ri_f^\infty < 1$. This conclusion by no means supports the idea of the critical gradient Richardson number. Indeed, $Ri_f$ is an internal parameter controlled by the turbulence in contrast to $Ri = (\beta \partial \Theta / \partial z)/(\partial \mathbf{U}/\partial z)^2$, which is an "external parameter" characterising the mean flow.

Recall that the key parameter characterising the vertical turbulent transports is the TKE of the vertical velocity fluctuations, $E_z = \tfrac{1}{2}\langle w^2 \rangle$, rather than the TKE as such. To determine $E_z$, we need to consider all three budget equations (10a) for the diagonal Reynolds stresses: $\tau_{11} = 2E_1 = 2E_x = \langle u^2 \rangle$, $\tau_{22} = 2E_2 = 2E_y = \langle v^2 \rangle$ and $\tau_{33} = 2E_3 = 2E_z = \langle w^2 \rangle$. In the steady state they become

$$\frac{E_i}{C_K t_T} = -\tau_{i3}\frac{\partial U_i}{\partial z} + \frac{1}{2}Q_{ii}, \quad i=1,2, \tag{27a}$$

$$\frac{E_z}{C_K t_T} = \frac{E_3}{C_K t_T} = \beta F_z + \frac{1}{2}Q_{33}. \tag{27b}$$

The sum of the pressure – velocity shear correlation terms, $\sum Q_{ii} = \sum \rho_0^{-1}\langle p\,\partial u_i/\partial x_i\rangle$, is zero because of the continuity equation: $\sum \partial u_i / \partial x_i = 0$. Hence they are neither productive nor dissipative and describe the conversion of the energy of "rich" component into the energy of "poorer" components.

To determine the diagonal terms, $Q_{11}$, $Q_{22}$ and $Q_{33}$, we generalize the familiar "return to isotropy" hypothesis as follows:

$$Q_{11} = -\frac{2C_r}{3C_K t_T}(3E_1 - E_K \Psi_1), \qquad Q_{22} = -\frac{2C_r}{3C_K t_T}(3E_2 - E_K \Psi_2),$$

$$Q_{33} = -\frac{2C_r}{3C_K t_T}(3E_3 - E_K \Psi_3). \tag{28}$$

Here, $C_r$ and $\Psi_i$ ($i=1,2,3$) are dimensionless empirical coefficients; $C_r$ accounts for the difference between the relaxation-time and the dissipation-time scales (as a first

---

[2] Admitting that $C_P$ and $C_K$ could differ, $Ri_f$ is proportional rather than equal to $E_P/E$ (see Zilitinkevich et al., 2007)

approximation, we take these two time scales proportional: $t_r \sim t_T$, so that $C_r = t_r/t_T =$ constant); $\Psi_i$ govern redistribution of TKE between the components. At $\Psi_i = 1$ the above formulae reduce to their original form proposed by Rotta (1951) and known to be a good approximation for neutrally stratified flows. In the stable stratification, we need to leave room for their possible stability dependence. As a first approximation we assume

$$\Psi_i = 1 + C_i \operatorname{Ri}_f, \quad i = 1,2,3, \qquad (29)$$

where $\operatorname{Ri}_f$ is the flux Richardson number, and $C_i$ are empirical constants. Their sum must be zero: $C_1 + C_2 + C_3 = 0$, to satisfy the condition $\sum Q_{ii} = 0$ (needed to guarantee that $E_K = E_1 + E_2 + E_3$). Linear functions of $\operatorname{Ri}_f$ on the r.h.s. of Eq. (29) are taken as simple approximations providing the only possible from the physical point of view, finite, non-zero limits: $\Psi_i = 1$ at $\operatorname{Ri} = 0$, and $\Psi_i \to 1 + C_i \operatorname{Ri}_f^\infty$ at $\operatorname{Ri} \to \infty$.

Because the energy exchange between the horizontal components of the TKE, $E_1$ and $E_2$, is not directly affected by the stable stratification, we take the first two energy-exchange constants equal: $C_1 = C_2$. Then, recalling the above condition: $C_1 + C_2 + C_3 = 0$, only one of the three constants is independent: $C_1 = C_2 = -\tfrac{1}{2} C_3$.

Equations (27)-(28) yield

$$E_i = \frac{C_r}{3(1+C_r)} E_K \Psi_i - \frac{C_K}{1+C_r} t_T \tau_{i3} \frac{\partial U_i}{\partial z}, \quad i=1,2, \qquad (30a)$$

$$E_z = \frac{C_r}{3(1+C_r)} E_K \Psi_3 + \frac{C_K}{1+C_r} t_T \beta F_z. \qquad (30b)$$

In the plain-parallel neutral boundary layer with $\mathbf{U} = (U,0,0)$, Eqs. (30a) and (30b) reduce to

$$\frac{E_x}{E_K} = \frac{3+C_r}{3(1+C_r)}, \qquad (31a)$$

$$\frac{E_y}{E_K} = \frac{E_z}{E_K} = \frac{C_r}{3(1+C_r)}. \qquad (31b)$$

Given the vertical component of the TKE, $E_z$, the turbulent dissipation time scale, $t_T = l_T E_K^{-1/2}$, can alternatively be expressed through the vertical turbulent length scale $l_z$:

$$t_T = \frac{l_z}{E_z^{1/2}}. \qquad (32)$$

Then eliminating $t_T$ from Eq. (25) for $E_K$ and Eq. (30b), and substituting Eq. (29) for $\Psi_3$ yields

$$E_z = \left[\frac{C_K C_r \Psi_3}{3(1+C_r)}\left(\Pi + \left(\frac{3}{C_r \Psi_3}+1\right)\beta F_z\right)l_z\right]^{2/3}, \tag{33a}$$

$$\Psi_3 = 1 + C_3 \operatorname{Ri}_f. \tag{33b}$$

This formulation reduces to the traditional return-to-isotropy formulation taking $C_3 = 0$.

To close the system, the horizontal components of the TKE, $E_x$ and $E_y$, are not required. We leave their discussion to a separate paper, in which our closure is extended to passive scalars and applied to the turbulent diffusion.

## 4.2. VERTICAL TURBULENT FLUXES OF MOMENTUM AND POTENTIAL TEMPERATURE

Of the non-diagonal Reynolds stresses we consider only those representing the vertical fluxes of momentum $\tau_{13} = \tau_{xz} = \langle uw \rangle$ and $\tau_{23} = \tau_{yz} = \langle vw \rangle$ needed to close the momentum equations (4)-(5) and determined by Eqs. (10b). In the steady state, using Eqs. (17)-(18) for the effective Reynolds-stress dissipation rate, they become

$$\tau_{i3} = -2\Psi_\tau E_z^{1/2} l_z \frac{\partial U_i}{\partial z}. \tag{34}$$

Of the three components of the potential-temperature flux, we consider only the vertical flux $F_3 = F_z$ needed to close the thermodynamic energy Equation (6) and determined by Eq. (9b). Taking $\beta E_\theta = (N^2/\beta) E_P = -C_K N^2 l_z F_z / E_z^{1/2}$ [after Eqs. (25) and (32)], the steady-state version of Eq. (9b) becomes

$$F_z = -\frac{2C_F E_z^{1/2} l_z}{1 + 2C_\theta C_F C_K (N l_z)^2 E_z^{-1}} \frac{\partial \Theta}{\partial z}. \tag{35}$$

Substituting here $N^2 = \beta \partial \Theta/\partial z$ shows that $F_z$ depends on $\partial \Theta/\partial z$ weaker then linearly and at $\partial \Theta/\partial z \to \infty$ tends to a finite limit:

$$F_{z,\max} = -\frac{E_z^{3/2}}{C_\theta C_K \beta l_z}. \tag{36}$$

It follows that $F_z$ in a turbulent flow cannot be considered as given external parameter. This conclusion is consistent with our reasoning (in Section 4.1) that the flux Richardson number $\mathrm{Ri}_f = -\beta F_z (\tau S)^{-1}$ is an internal parameter of turbulence that cannot be arbitrarily prescribed. According to Eq. (36), the maximal value of the buoyancy flux in very stable stratification, $\beta F_z$, is proportional to the dissipation rate, $E_z^{3/2}(C_K l_z)^{-1}$, of the energy of vertical velocity fluctuations[3].

Equations (34) and (35) allow determining the eddy viscosity and conductivity:

$$K_M \equiv \frac{-\tau_{i3}}{\partial U_i / \partial z} = 2\Psi_\tau E_z^{1/2} l_z, \qquad (37a)$$

$$K_H \equiv \frac{-F_z}{\partial \Theta / \partial z} = \frac{2 C_F E_z^{1/2} l_z}{1 + 2 C_\theta C_F C_K (N l_z)^2 E_z^{-1}}. \qquad (37b)$$

Thus the Kolmogorov's closure hypothesis applied to the effective Reynolds-stress dissipation rate, Eqs. (17)-(18), leads to the eddy-viscosity formulation, Eq. (37a), basically similar to the traditional formulation, Eq. (1), whereas Eq. (37b) for the eddy conductivity differs essentially from this formulation.

It may appear that our derivation of Eq. (37a) essentially depends on the hypothetical concept of the effective dissipation rare, Eqs. (17)-(18). In fact we employ this trick merely for reader's convenience, to avoid too complicated derivations. Principally the same result, namely the down-gradient momentum-flux formulation equivalent to Eqs. (34) and (37a), follows from analyses of the budget equations for the Reynolds stresses in the k-space using the familiar "$\tau$-approximation" (e.g. Elperin et al., 2002, 2006).

Recall that $\Psi_\tau$ is a dimensionless, non-zero, limited coefficient that could only monotonically depend on the static stability [see Eqs. (17)-(18) and their discussion in Section 2]. We approximate its stability dependencies by a linear function of the flux Richardson number, $\mathrm{Ri}_f$:

$$\Psi_\tau = C_{\tau 1} + C_{\tau 2} \mathrm{Ri}_f, \qquad (38)$$

---

[3] A principally similar analysis of the budget equation for $F_z$ has been performed by Cheng et al. (2002). Their Eq. (15i) implies the same maximal value of $F_z$ as our Eq. (36). It worth noticing that Eq. (35) imposes an upper limit on the downward heat flux in the deep ocean (known to be a controlling factor of the rate of global warming).

where $C_{\tau 1}$ and $C_{\tau 2}$ are dimensionless constants to be determined empirically. Equation (38) provides the only physically meaningful, finite, non-zero limits: $\Psi_\tau = C_{\tau 1}$ at Ri = 0, and $\Psi_\tau \to C_{\tau 1} + C_{\tau 2} \mathrm{Ri}_f^\infty$ at Ri $\to \infty$ [cf. our argument in support of Eq. (29)].

## 4.3. TURBULENT PRANDTL NUMBER AND OTHER DIMENSIONLESS PARAMETERS

The system of Equations (33)-(35), although unclosed until we determine the vertical turbulent length scale $l_z$, reveals a "partial invariance" with respect to $l_z$ and allows determining the turbulent Prandtl number, $\mathrm{Pr}_T$, the flux Richardson number, $\mathrm{Ri}_f$, and some other dimensionless characteristics of turbulence as universal functions of the gradient Richardson number, Ri. Certainly such universality is relevant only to the steady-state homogeneous regime. In non-steady, heterogeneous regimes, all these characteristics will no longer be single-valued functions of Ri.

Recalling that $\Pi = K_M S^2$ and $\mathrm{Ri}_f \equiv -\beta F_z / \Pi$, Eqs. (33) and (37) give

$$\frac{E_z}{(Sl_z)^2} = \Psi(\mathrm{Ri}_f) \equiv \frac{2C_K C_r \Psi_3 \Psi_\tau}{3(1+C_r)} \left[1 - \left(\frac{3}{C_r \Psi_3} + 1\right) \mathrm{Ri}_f \right], \tag{39}$$

where $\Psi_3$ and $\Psi_\tau$ are linear functions of $\mathrm{Ri}_f$ given by Eqs. (33b) and (38). Then dividing $K_M$ [determined after Eq. (37b)] by $K_H$ [determined after Eq. (38)] and expressing $E_z$ through Eq. (39) yields surprisingly simple expressions:

$$\mathrm{Pr}_T \equiv \frac{K_M}{K_H} = \frac{\mathrm{Ri}}{\mathrm{Ri}_f} = \frac{\Psi_\tau}{C_F} + \frac{3(1+C_r)C_\theta}{C_r \Psi_3} \mathrm{Ri} \left[1 - \left(\frac{3}{C_r \Psi_3} + 1\right) \mathrm{Ri}_f \right]^{-1}, \tag{40}$$

and

$$\frac{1}{\mathrm{Ri}} = \frac{C_F \Psi_\tau^{-1}}{\mathrm{Ri}_f} - \frac{3C_F(1+C_r)C_\theta \Psi_\tau^{-1}}{C_r \Psi_3 (1-\mathrm{Ri}_f) - 3\mathrm{Ri}_f}, \tag{41}$$

which do not include $l_z$. Equation (41) together with Eqs. (33b) and (38) specify Ri as a single-valued, monotonically increasing function of $\mathrm{Ri}_f$ determined in the interval $0 < \mathrm{Ri}_f < \mathrm{Ri}_f^\infty$, where $\mathrm{Ri}_f^\infty$ is given by Eq. (45). Therefore, the inverse function, namely,

$$\mathrm{Ri}_f = \Phi(\mathrm{Ri}) \tag{42}$$

is a monotonically increasing function of $\text{Ri}_f$, changing from 0 at $\text{Ri}_f = 0$ to infinity at $\text{Ri}_f = \text{Ri}_f^\infty$.

According to the above equations, the Ri-dependencies of $\text{Ri}_f$ and $\text{Pr}_T$ (which also monotonically increases with increasing Ri) are characterised by the following asymptotic limits:

$$\text{Pr}_T \approx \frac{\Psi_\tau^{(0)}}{C_F} + \left( \frac{3C_\theta (1+C_r)}{C_r} + \frac{C_{\tau 2}}{C_F} \right) \text{Ri} \rightarrow \text{Pr}_T^{(0)} = \frac{\Psi_\tau^{(0)}}{C_F}, \tag{43a}$$

$$\text{Ri}_f \approx \frac{C_F}{\Psi_\tau^{(0)}} \text{Ri} \quad \text{at} \quad \text{Ri} \ll 1, \tag{43b}$$

$$\text{Pr}_T \approx \frac{1}{\text{Ri}_f^\infty} \text{Ri}, \tag{44a}$$

$$\text{Ri}_f \rightarrow \text{Ri}_f^\infty \quad \text{at} \quad \text{Ri} \gg 1, \tag{44b}$$

where

$$\text{Ri}_f^\infty = \frac{C_r \Psi_3^\infty}{C_r \Psi_3^\infty + 3[1 + C_\theta (1 + C_r)]}, \tag{45}$$

and the superscripts "(0)" and "∞" mean "at Ri=0" and "at Ri→∞", respectively.

Equations (33)-(35) allow determining, besides $\text{Pr}_T$, some other dimensionless parameters, in particular, the vertical anisotropy of turbulence:

$$A_z \equiv \frac{E_z}{E_K} = \frac{C_r \Psi_3}{3(1+C_r)} \left[ 1 - \left( \frac{3}{C_r \Psi_3} + 1 \right) \text{Ri}_f \right] (1 - \text{Ri}_f)^{-1}, \tag{46}$$

the squared ratio of the turbulent flux of momentum to the TKE (characterising the correlation between vertical and horizontal velocity fluctuations):

$$\left( \frac{\tau}{E_K} \right)^2 = \frac{2\Psi_\tau A_z}{C_K (1 - \text{Ri}_f)}, \tag{47}$$

and the ratio of the squared vertical flux of potential temperature to the product of the TKE and the "energy" of the potential temperature fluctuation fluctuations:

$$\frac{F_z^2}{E_K E_\theta} = \frac{2\Psi_\tau A_z}{C_K \Pr_T}. \tag{48}$$

Equations (46)-(48) in combination with Eq. (41) determine the Ri-dependencies of $A_z$, $\tau^2 E_K^{-2}$ and $F_z^2(E_K E_\theta)^{-2}$, characterised by the following asymptotic limits:

$$A_z \to A_z^{(0)} = \frac{C_r}{3(1+C_r)}, \tag{49a}$$

$$\left(\frac{\tau}{E_K}\right)^2 \to \frac{2\Psi_\tau^{(0)} A_z^{(0)}}{C_K}, \tag{49b}$$

$$\frac{F_z^2}{E_K E_\theta} \to \frac{2 C_F A_z^{(0)}}{C_K} \quad \text{at Ri} \ll 1, \tag{49c}$$

$$A_z \to A_z^\infty = \frac{C_r \Psi_3^\infty}{3(1+C_r)}\left[1 - \left(\frac{3}{C_r \Psi_3^\infty}+1\right)\mathrm{Ri}_f^\infty\right](1-\mathrm{Ri}_f^\infty)^{-1}, \tag{50a}$$

$$\left(\frac{\tau}{E_K}\right)^2 \to \frac{2\Psi_\tau^\infty A_z^\infty}{C_K(1-\mathrm{Ri}_f^\infty)}, \tag{50b}$$

$$\frac{F_z^2}{E_K E_\theta} \to \frac{2\Psi_\tau^\infty A_z^\infty}{C_K \Pr_T^\infty} \quad \text{at Ri} \gg 1. \tag{50c}$$

Recall that the turbulent velocity scale in Eqs. (34)-(37) is $\sqrt{E_z}$ rather than $\sqrt{E_K}$. However, in a number of currently used turbulence closure models the stability dependence of $A_z = E_z/E_K$ is neglected and $\sqrt{E_K}$ is taken as an ultimate velocity scale characterising the vertical turbulent transports, without serious theoretical or experimental grounds. On the contrary, Eq. (48) implies an essential Ri-dependence of $A_z$, in agreement with currently available data [see Mauritsen and Svensson (2007) and our data analysis in Section 6 below].

## 4.4. VERTICAL TURBULENT LENGTH SCALE

The basic factors that impose limits on the vertical turbulent length scale, $l_z$, in geophysical flows are the height over the surface (the geometric limit) and the stable stratification.

In neutral stratification, $l_z$ is restricted by the geometric limit[4]:

$$l_z \sim z. \qquad (51)$$

For the stable stratification limit, different formulations have been proposed. Consider, first, the Monin and Obukhov (1954) length scale widely used in boundary-layer meteorology:

$$L \equiv \frac{\tau^{3/2}}{-\beta F_z} = \frac{\tau^{1/2}}{S \, \mathrm{Ri}_f}. \qquad (52)$$

Equations (34), (39) and (52) give

$$l_z = \frac{\mathrm{Ri}_f}{(2\Psi_\tau)^{1/2} \Psi^{1/4}} L. \qquad (53)$$

The r.h.s. of this formula specifies a universal function of $\mathrm{Ri}_f$. Hence any interpolation formula for $l_z$ linking the limits $l_z \sim z$ and $l_z \sim L$ should have the form:

$$l_z = z \, \Psi_l(\mathrm{Ri}_f), \qquad (54)$$

where $\Psi_l$ is a function of $\mathrm{Ri}_f$.

Well known alternatives to $L$ are the Ozmidov scale: $\varepsilon_K^{1/2} N^{3/2}$ (Ozmidov, 1990), the local energy balance scale: $E_z^{1/2} N^{-1}$ (e.g., Table 3 in Cuxart et al., 2006), the shear sheltering scale: $E_z^{1/2} S^{-1}$ (Hunt et al., 1985, 1988); the list could be extended. Using our local closure equations (Sections 4.1-4.3) the ratio of each of these scales to $L$ can be expressed through corresponding function of $\mathrm{Ri}_f$. Hence any interpolation linking the neutral stratification limit, $l_z \sim z$, with all the above limits will still have the same form as Eq. (54).

It follows that Eq. (54) represents a general formulation for the vertical turbulent length scale in the steady-state, homogeneous, stably stratified flows. In other words, the stability dependence of $l_z$ is fully characterised by the universal function $\Psi_l(\mathrm{Ri}_f)$. The latter should satisfy the following physical requirements. In the neutral stratification it achieves the maximal value: $\Psi_l(0) = 1$ [the omitted empirical constant combines with the coefficients $C_K = C_P$, $C_F$ and $\Psi_\tau$ in Eqs. (16) and (18)]. With increasing $\mathrm{Ri}_f$ it should monotonically decrease. And at $\text{Ri}_f \to \text{Ri}_f^\infty$ it should tend to zero [otherwise Eq. (33) would give $E_z > 0$ at $\text{Ri}_f \to \text{Ri}_f^\infty$, which is physically senseless].

---

[4] In rotating fluids, the direct effect of the angular velocity, $\Omega$, on turbulent eddies is characterised by the rotational limit, $E_z^{1/2}/\Omega$. In geophysical, stably stratified flows it plays only a secondary role. We leave its discussion for future work.

We propose a simple approximation satisfying these requirements: $\Psi_l = \left(1 - \text{Ri}_f / \text{Ri}_f^\infty\right)^n$, where $n$ is a positive constant. With empirical value of $n = 4/3$ (see next Section) it gives

$$l_z = z \left(1 - \frac{\text{Ri}_f}{\text{Ri}_f^\infty}\right)^{4/3}. \tag{55}$$

Certainly, in non-steady, heterogeneous regimes $l_z$ should be determined through a prognostic equation accounting for its advection and temporal evolution.

## 5. Empirical verification of the local model

To determine empirical dimensionless constants $C_r$, $C_K$, $C_F$, $C_\theta$, $C_{\tau 1}$, $C_{\tau 2}$, $C_3$ and $n$ we compare results from the local closure model given in Section 4 with experimental, large-eddy simulation (LES) and direct numerical simulation (DNS) data.

Recall that the local model is applied to the homogeneous turbulence and does not include transports of turbulent energies and turbulent fluxes. At the same time practically all currently available data represent vertically (in a number of cases, vertically and horizontally) heterogeneous flows, in which the above transports are more or less pronounced. In these conditions, fundamental dimensionless parameters of turbulence, such as $\text{Pr}_T$, $\text{Ri}_f$, $(\tau / E_K)^2$, $F_z^2 /(E_K E_\theta)$ and $A_z$, can be considered as universal functions of Ri, if et all, only approximately. Anyhow Mauritsen and Svensson (2007) and Zilitinkevich et al. (2007) have demonstrated quite reasonable Ri-dependencies of the above parameters based on data sets from several recent field campaigns and numerical simulations. To reduce inevitable deviations from universality and to more accurately determine empirical constants, we now more carefully select data and rule out those representing strongly heterogeneous regimes.

Figures 1a,b show turbulent Prandtl number, $\text{Pr}_T$, and flux Richardson number, $\text{Ri}_f = \text{Ri} / \text{Pr}_T$, versus gradient Richardson number, Ri. They demonstrate reasonable agreement between data from atmospheric and laboratory experiments, LES and DNS. Data for $\text{Ri} \to 0$ in Figure 1 are consistent with the commonly accepted empirical estimate of $\text{Pr}_T^{(0)} \equiv \text{Pr}_T |_{\text{Ri} \to \infty} = 0.8$ [see data collected by Churchill (2002) and Foken (2006) and theoretical analysis of Elperin et al. (1996a)]. Figure 1b clearly demonstrates that $\text{Ri}_f$ at large Ri levels off, and allows estimating its limiting value: $\text{Ri}_f^\infty = 0.2$.

Figure 2 shows Ri-dependences of the dimensionless turbulent fluxes: (a) $\hat{\tau}^2 \equiv (\tau/E_K)^2$ and (b) $F_z^2/(E_K E_\theta)$. It has been recognised long ago [see, e.g., Sections 5.3 and 8.5 in Monin and Yaglom (1971)] that in neutral stratification atmospheric data give more variable and generally smaller values of these ratios than lab experiments. This is not surprising because measured values of the TKE, $E_K$, in the atmosphere factually include low-frequency velocity fluctuations caused by the interaction of the air-flow with the surface heterogeneities. These low-frequency fluctuations should not be confused with the shear-generated turbulence. Therefore, to validate our turbulence closure model it is only natural to use data on $\hat{\tau}^2$ obtained from lab experiments and/or numerical simulations. Relying on this kind of data presented in Figure 2a, we obtain $(\tau/E_K)^{(0)} = 0.326$ for Ri<<1; and $(\tau/E_K)^\infty = 0.18$ for Ri>>1 [the superscripts "(0)" and "∞" mean "at Ri=0" and "at Ri→∞"]. These estimates are consistent with the conditions $(\hat{\tau}^2)^{(0)}/(\hat{F}_z^2)^{(0)} = \Pr_T^{(0)} = 0.8$, and $(\hat{F}_z^2)^\infty = 0$ followed from our Equations (47)-(48). Furthermore, Figure 3 showing re-normalised fluxes: (a) $\hat{\tau}^2/(\hat{\tau}^2)^{(0)}$ and (b) $\hat{F}_z^2/(\hat{F}_z^2)^{(0)}$ as dependent on Ri, reveal essential similarity in the shape of these dependences after atmospheric, laboratory and LES data, and by these means provide additional support to our analysis.

Data on the vertical anisotropy of turbulence, $A_z = E_z/E_K$, are shown in Figure 4. They are most ambiguous and need to be analysed carefully. For neutral stratification, we adopt the estimate of $A_z^{(0)} = 0.25$ based on precise laboratory experiments (Agrawal et al., 2004) and DNS (Moser et al., 1999) – now commonly accepted and shown to be consistent with independent data on the wall-layer turbulence (L'vov et al., 2006). Atmospheric data both new and prior (e.g. those shown in Figure 75 in the textbook of Monin and Yaglom (1971) give smaller values of $A_z^{(0)}$; but, as already mentioned, they overestimate the horizontal TKE and therefore underestimate $A_z$, especially in neutral stratification, due to meandering of atmospheric boundary-layer flows caused by non-uniform features of the earth's surface (hills, houses, groups of trees, etc.). At the same time, very large values of Ri in currently available experiments and numerical simulations are relevant to turbulent flows above the boundary layer, where the TKE of local origin (controlled by local Ri) is often small compared to the TKE transported from the lower, strong-shear layers. It is not surprising that the spread of data on $A_z$ versus Ri is quite large. Anyhow atmospheric data characterise $A_z$ as a monotonically decreasing function of Ri and allow at least approximately estimating its lover limit: $A_z^\infty = 0.075$.

Below we use the above estimates of $A_z^{(0)}$, $(\tau^2 E_K^{-2})_{\text{Ri}=0}$, $\Pr_T^{(0)}$, $\text{Ri}_f^\infty$, $A_z^\infty$, and $(\tau^2 E_K^{-2})_{\text{Ri}=\infty}$ to determine our empirical constants.

We start with data for neutral stratification. The empirical estimate of $A_z^{(0)} = 0.25$ yields

$$C_r = 3A_z^{(0)}(1 - 3A_z^{(0)})^{-1} = 3. \tag{56}$$

Then we combine Eq. (25) for $E_K$ with Eq. (32) for $t_T$ and consider the logarithmic boundary layer, in which $l_z = z$, $\tau = \tau|_{z=0} \equiv u_*^2$ and $S = u_*(kz)^{-1}$ ($u_*$ is the friction velocity and $k$ is the von Karman constant) to obtain

$$C_K = k(A_z^{(0)})^{1/2}\left(\frac{E_K}{\tau}\right)^{3/2}_{Ri=0} = 1.08. \tag{57}$$

This estimate is based on the well-determined empirical value of $k \approx 0.4$, and the above values of $(\tau/E_K)^{(0)} = 0.326$ and $A_z^{(0)} = 0.25$. Then taking $C_K = 1.08$ and $\Pr_T^{(0)} = 0.8$, Eqs. (43a) and (47) give

$$C_{\tau 1} = \frac{C_K}{2A_z^{(0)}}\left(\frac{E_K}{\tau}\right)^{-2}_{Ri=0} = 0.228, \tag{58}$$

$$C_F = C_{\tau 1}/\Pr_T^{(0)} = 0.285. \tag{59}$$

Taking $C_r = 3$, $A_z^\infty = 0.075$ and $Ri_f^\infty = 0.2$, Eq. (46) gives

$$\Psi_3^\infty = \frac{A_z^\infty}{A_z^{(0)}} + \frac{3Ri_f^\infty}{C_r(1-Ri_f^\infty)} = 0.55; \quad C_3 = \frac{1}{Ri_f^\infty}(\Psi_3^\infty - 1) = -2.25. \tag{60}$$

The constants $C_1$ and $C_2$ control only the energy exchange between the horizontal velocity components and do not affect any other aspects of our closure model. Taking them equal (from the symmetry reasons) and recalling that $C_1 + C_2 + C_3 = 0$ gives

$$C_1 = C_2 = -\frac{1}{2}C_3 = 1.125. \tag{61}$$

Taking $C_K = 1.08$, $Ri_f^\infty = 0.2$, $A_z^\infty = 0.075$ and $(\tau/E_K)^\infty = 0.18$, Eq. (50b) gives

$$\Psi_\tau^\infty = \frac{C_K[(\tau/E_K)^\infty]^2(1-Ri_f^\infty)}{2A_z^\infty} = 0.187; \quad C_{\tau 2} = \frac{1}{Ri_f^\infty}(\Psi_\tau^\infty - C_{\tau 1}) = -0.208 \tag{62}$$

Then $C_\theta$ is determined from Eq. (45) taken at the strong stability limit:

$$C_\theta = \frac{1}{1+C_r}\left[\frac{C_r\Psi_3^\infty}{3}\left(\frac{1}{Ri_f^\infty} - 1\right) - 1\right] = 0.3. \tag{63}$$

Given the above values of dimensionless constants, the function $\text{Ri}_f = \Phi(\text{Ri})$ calculated after Eq. (41)-(42) is shown in Figure 1b by solid line. For practical use we propose its explicit approximation (within 5 % accuracy):

$$\text{Ri}_f = \Phi(\text{Ri}) \approx 1.25 \, \text{Ri} \frac{(1+36\,\text{Ri})^{1.7}}{(1+19\,\text{Ri})^{2.7}}. \tag{64}$$

The latter is shown in Figure 1b by dashed line.

In the above estimates we did not use data on the dimensionless heat flux $\hat{F}_z^2 \equiv F_z^2/(E_K E_\theta)$ shown in Figures 2b and 3b. Quite good correspondence between data and theoretical curves in these figures serves an empirical confirmation to our model.

The last empirical constant to be determined is the exponent $n$ in Eq. (55). We eliminate $l_z$ from Eqs. (53) and (54) to obtain

$$\frac{z}{L} = \frac{\text{Ri}_f}{(2\Psi_\tau)^{1/2} \Psi^{1/4} \Psi_l}, \tag{65}$$

where $\Psi_\tau$ and $\Psi$ are functions of $\text{Ri}_f$ specified by Eqs. (38) and (39). Given the dependence $\Psi_l(\text{Ri}_f)$, the pair of Equation (65) and Eq. (41) determine $\text{Ri}_f$ and Ri as single-valued functions of $z/L$. And *vice versa*, given, e.g., the dependence Ri($z/L$), they allows determining $\Psi_l$ as a single-valued function of $\text{Ri}_f$.

We apply this kind of analysis to deduce $\Psi_l(\text{Ri}_f)$ from the empirical dependence of Ri on $z/L$ obtained by Zilitinkevich and Esau (2007) using LES DATABASE64 (Beare et al., 2006; Esau and Zilitinkevich, 2006) and data from the field campaign SHEBA (Uttal et al., 2002). In Figure 5, we present the above LES data together with our approximation based on Eq. (55). The exponent $n = 4/3$ is just obtained from the best fit of the theoretical curve to all data.

Strictly speaking, the local, algebraic closure model under consideration is applicable only to the homogeneous flows, in particular, to the nocturnal stable atmospheric boundary layer (ABL) of depth $h$, where non-local vertical turbulent transports play comparatively minor roles, whereas $\tau$ and $F_z$ are reasonably accurately represented by universal functions of $z/h$ (see, e.g., Figure 1 in Zilitinkevich and Esau, 2005); or with more confidence to its lower 10%, the so-called surface layer, where $\tau$ and $F_z$ can be taken depth-constant: $\tau \approx \tau|_{z=0} = u_*^2$ and $F_z \approx F_z|_{z=0} = F_*$.

Now recall that Eqs. (64) and (65) determine $\text{Ri}_f$ and Ri as single-valued functions of $z/L$:

$$\mathrm{Ri}_f = \Phi_{\mathrm{Rif}}\left(\frac{z}{L}\right), \tag{66a}$$

$$\mathrm{Ri} = \Phi_{\mathrm{Ri}}\left(\frac{z}{L}\right). \tag{66b}$$

It follows that our model – as applied to the steady-state, homogeneous regime in the surface layer – is consistent with the Monin and Obukhov (1954) similarity theory. Given $\tau$ and $F_z$, it allows determining *z/L*-dependences of all dimensionless parameters considered above, as well as the familiar similarity-theory functions specifying mean velocity and temperature profiles:

$$\Phi_M \equiv \frac{kz}{\tau^{1/2}}\frac{\partial U}{\partial z} \equiv \frac{k}{\mathrm{Ri}_f}\frac{z}{L} = \frac{k}{\Phi_{\mathrm{Rif}}(z/L)}\frac{z}{L}, \tag{67a}$$

$$\Phi_H \equiv \frac{k_T z \tau^{1/2}}{-F_z}\frac{\partial \Theta}{\partial z} \equiv k_T \frac{\mathrm{Pr}_T}{\mathrm{Ri}_f}\frac{z}{L} \equiv k_T \frac{\mathrm{Ri}}{\mathrm{Ri}_f^2}\frac{z}{L} = k_T \frac{\Phi_{\mathrm{Ri}}(z/L)}{\Phi_{\mathrm{Rif}}^2(z/L)}\frac{z}{L}, \tag{67b}$$

where $k$ is the von Karman constant expressed through our constants by Eq. (57) and $k_T = k/\mathrm{Pr}_T^{(0)}$. At Ri<<1, Eq. (66a,b) reduce to $\mathrm{Ri}_f \approx kz/L$ and $\mathrm{Ri} \approx \mathrm{Pr}_T^{(0)} kz/L$; and Eq. (67a,b) reduce to the familiar wall-layer formulation.

$\Phi_M(z/L)$ and $\Phi_H(z/L)$ calculated after our model are shown in Figure 6 together with LES data from Zilitinkevich and Esau (2007).

In contrast to the commonly accepted idea that both $\Phi_M$ and $\Phi_H$ depend on *z/L* linearly, LES data and our solution show different asymptotic behaviours, namely, linear for $\Phi_M$ and stronger than linear for $\Phi_H$. This result deserves emphasising. Indeed, the traditional formulation: $\Phi_M, \Phi_H \sim z/L$ at *z/L*>>1 implies that $\mathrm{Pr}_T$ with increasing *z/L* levels off (rather than increases) and, as a consequence, that the surface-layer turbulence decays when Ri exceeds critical value, Ri$_c$~0.25. However, as demonstrated in Sections 1 and 3 this conclusion is erroneous (see more detailed discussion in Zilitinkevich et al., 2007).

Recall that the linear dependences: $\Phi_M \sim \Phi_H \sim$ *z/L* were traditionally derived from the heuristic "*z*-less stratification" concept, which postulates that the distance from the surface, *z*, drops out from the set of parameters characterising the vertical turbulent length scale in sufficiently strong static stability (*z/L* >> 1). Ruling out this concept, the linear asymptote for $\Phi_H$ loses grounds; but for $\Phi_M$ is holds true. Indeed the existence of finite upper limit for the flux Richardson number: $\mathrm{Ri}_f \to \mathrm{Ri}_f^\infty$ at *z/L*→∞ immediately yields the asymptotic formula:

$$\Phi_M \approx C_U \frac{z}{L} \quad \text{at} \quad \frac{z}{L} \gg 1, \quad \text{where} \quad C_U = (\text{Ri}_f^\infty)^{-1} \approx 5. \tag{68}$$

We emphasise that the algebraic closure model presented in Section 4 is applicable only to the homogeneous turbulence regimes. Therefore it probably serves as a reasonable approximation for the nocturnal ABLs separated from the free flow by the neutrally stratified residual layers, but not for the conventionally neutral and the long-lived stable ABLs, which develop against the stably stratified free flow and exhibit essentially non-local features such as the distant effect of the free-flow stability on the surface-layer turbulence (see Zilitinkevich, 2002; Zilitinkevich and Esau, 2005). To reproduce these types of the ABL realistically, an adequate turbulence closure model should take into account the non-local transports.

# Conclusions

The architecture of the most widely used turbulence closure models for neutrally and stably stratified geophysical flows follows Kolmogorov (1941): vertical turbulent fluxes are assumed to be down-gradient; the turbulent exchange coefficients, namely, the eddy viscosity, $K_M$, conductivity, $K_H$, and diffusivity, $K_D$, are taken proportional to the turbulent length scale, $l_T$, and the turbulent velocity scale, $u_T$, in its turn taken proportional to the square root of the TKE, $E_K^{1/2}$, so that $K_{\{M,H,D\}} \sim E_K^{1/2} l_T$; and $E_K$ is determined solely from the TKE budget equation. Kolmogorov has designed this formulation for the neutral stratification, where it provides quite good approximation. However, when applied to essentially stable stratification it predicts that the TKE decays at Richardson numbers exceeding critical value, $\text{Ri}_c$ (close to 0.25), which contradicts experimental evidence. To avoid this drawback, modern closure models modify the original Kolmogorov's formulation taking $K_{\{M,H,D\}} = f_{\{M,H,D\}}(\text{Ri}) E_K^{1/2} l_z$, where stability functions $f_{\{M,H,D\}}(\text{Ri})$ are determined either theoretically or empirically. Given these functions, it remains to determine $l_T$ and then, to the first sight, the closure problem is solved.

Such conclusion is premature. The concepts of the down-gradient turbulent transport and the turbulent exchange coefficients, as well as the relationships $K_{\{M,H,D\}} = f_{\{M,H,D\}}(\text{Ri}) E_K^{1/2} l_T$ are consistent with the flux-budget equations only in comparatively simple particular cases relevant to the homogeneous regime of turbulence. Only in these cases the turbulent exchange coefficients can be rigorously defined, in contrast to turbulent fluxes that represent clearly defined, measurable parameters, governed by the flux-budget equations. It is therefore preferable to rely on the flux-budget equations rather than to formulate hypotheses about virtual exchange coefficients.

Furthermore, the TKE budget equation does not fully characterise turbulent energy transformations, not to mention that the vertical turbulent transports are controlled by the energy of vertical velocity fluctuations, $E_z$, rather than $E_K$.

We do not follow the above "main stream". Instead of the sole use of the TKE budget equation, we employ the budget equations for turbulent potential energy (TPE) and turbulent total energy (TTE = TKE + TPE), which guarantees maintaining of turbulence by the velocity shear in any stratification.

Furthermore, we do not accept *a priori* the concept of down-gradient turbulent transports (implying universal existence of turbulent exchange coefficients). Instead, we use the budget equations for key turbulent fluxes and derive (rather than postulate) formulations for the exchange coefficients, when it is physically grounded – in the steady-state homogeneous regime.

In the budget equation for the vertical flux of potential temperature we take into account a crucially important mechanism: generation of the counter-gradient flux due to the buoyancy effect of potential-temperature fluctuations (compensated but only partially by the correlation between the potential-temperature and the pressure-gradient fluctuations). We show that this is just the mechanism responsible for the principle difference between the heat- and the momentum-transfer.

To determine the energy of the vertical velocity fluctuations, we modify the traditional return-to-isotropy formulation accounting for the effect of stratification on the redistribution of the TKE between horizontal and vertical velocity components.

In this paper we derive the simplest, algebraic version an energetically consistent closure model for the steady-state, homogeneous regime, and verify it against available experimental, LES and DNS data.

As seen from Figures 1-4 showing Ri-dependences of the turbulent Prandtl number, $Pr_T = K_M / K_H$, flux Richardson number, $Ri_f$, dimensionless turbulent fluxes, $(\tau / E_K)^2$ and $F_z^2 (E_K E_\theta)^{-1}$, and anisotropy of turbulence, $A_z = E_z / E_K$, our model as well as the majority of data disclose two essentially different regimes of turbulence separated by a comparatively narrow interval of Ri around a threshold value of Ri ~ 0.25 (shown in the figures by the vertical dashed lines). On both sides of the transition interval, 0.1< Ri <1, the ratios $(\tau / E_K)^2$ and $F_z^2 (E_K E_\theta)^{-1}$ approach plateaus corresponding to the very high efficiency of the turbulent transfer at Ri < 0.1, and to the strongly different efficiencies of the momentum transfer (still pronounced) and the heat transfer (very weak) at Ri > 1.

It is hardly accidental that the above threshold coincides with the critical Richardson number, $Ri_c$, derived from the classical perturbation analyses. The latter have demonstrated that the infinitesimal perturbations grow exponentially at Ri < $Ri_c$ but do not grow at Ri > $Ri_c$ when, as we understand now, the onset of turbulent events requires finite perturbations [see Zilitinkevich et al. (2007)]. It follows that the transition interval, 0.1 < Ri <1, indeed separates two essentially different regimes: strong turbulence at Ri < 0.1 and weak turbulence at Ri > 1; but not the turbulent and the laminar regimes as classics assumed.

This paper starts rather than completes the development of consistent and practically useful turbulence closure models based on minimal sets of equations, indispensably including the TTE budget equation and free of the critical Richardson number. Two other new works follow this approach: Mauritsen et al. (2007) have developed a simpler closure model employing the TTE budget equation and empirical Ri-dependences of $(\tau/E_K)^2$ and $F_z^2 (E_K E_\theta)^{-1}$ (similar to those shown in our Figures 2-3); L'vov et al. (2007) have performed detailed analyses of the budget equations for the Reynolds stresses in the turbulent boundary layer (relevant to the strong turbulence regime) considering the dissipative effect of the horizontal heat flux explicitly, in contrast to our "effective-dissipation approximation".

As already mentioned, the present paper limits to the local, algebraic closure model applicable to the steady-state, homogeneous turbulence regime. Its generalised version, based on the same physical analyses but accounting for the third-order transports ($\Phi_K$, $\Phi_P$, $\Phi_F$ and $\Phi_{\{1,2\}}^{(\tau)}$) will be given in forthcoming papers.

Our data analysis gives only a plausible first verification rather than comprehensive validation of the proposed model. Special efforts are needed to extend our data analysis using additional field, laboratory and numerically-simulated data (e.g., Rohr et al., 1988; Shih et al., 2000). In future work, particular attention should also be paid to direct verification of our approximations, such as those for the term $\rho_0^{-1}\langle\theta\,\partial p/\partial z\rangle$ taken proportional to $\beta\langle\theta^2\rangle$ in Eq. (9b), and for the term $\varepsilon_{i3(\text{eff})} \equiv \varepsilon_{i3}^{(\tau)} - \beta F_i - Q_{i3}$ taken proportional to $\tau_{i3}/t_T$ in Eq. (10b).

In its present state our closure model does not account for the vertical transports due to internal waves. The dual nature of fluctuations representing both turbulence and waves in stratified flows was emphasized, e.g., by Jacobitz et al. (2005). The role of waves and necessity of their inclusion in the context of turbulence closure models has been discussed, e.g., by Jin et al. (2003) and Baumert and Peters (2005). Direct account for the wave-driven transports of momentum and both kinetic and potential energies is also in our plans.

# Acknowledgements

We thank for discussions Victor L'vov, Vittorio Canuto, Igor Esau, Thorsten Mauritsen and Gunilla Svensson. This work has been supported by EU Marie Curie Chair Project MEXC-CT-2003-509742, EU Project FUMAPEX EVK4-CT-2002-00097, ARO Project W911NF-05-1-0055, Carl-Gustaf Rossby International Meteorological Institute in Stockholm, German-Israeli Project Cooperation (DIP) administrated by the Federal

# Appendix A: The pressure term in the budget equation for the turbulent flux of potential temperature

The approximation used in Section 2:

$$\beta \langle \theta^2 \rangle + \frac{1}{\rho_0} \langle \theta \frac{\partial}{\partial z} p \rangle = C_\theta \beta \langle \theta^2 \rangle \tag{A1}$$

with $C_\theta$ = constant < 1 is justified as follows. Taking the divergence of the momentum equation:

$$\frac{1}{\rho_0} \Delta p = -\beta \frac{\partial}{\partial z} \theta, \tag{A2}$$

and applying to Eq. (A2) the inverse Laplacian yields

$$\frac{1}{\rho_0} p = \beta \Delta^{-1} \left( \frac{\partial \theta}{\partial z} \right), \quad \text{and} \quad \frac{1}{\rho_0} \langle \theta \frac{\partial}{\partial z} p \rangle = -\beta \langle \theta \Delta^{-1} \frac{\partial^2}{\partial z^2} \theta \rangle. \tag{A3}$$

Then we employ the scaling estimate:

$$\frac{\langle \theta \Delta^{-1} \left( \frac{\partial^2 \theta}{\partial z^2} \right) \rangle}{\langle \theta^2 \rangle} \approx (1 + \alpha^{-1}) \left( 1 - \frac{\arctan \sqrt{\alpha}}{\sqrt{\alpha}} \right), \tag{A4}$$

where $\alpha = l_\perp^2 / l_z^2 - 1$, $l_z$ and $l_\perp$ are the correlation lengths of the correlation function $\langle \theta(t, \mathbf{x}_1) \theta(t, \mathbf{x}_2) \rangle$ in the vertical and the horizontal directions.

Eqs. (A3) and (A4) yield

$$\frac{\frac{1}{\rho_0} \langle \theta \frac{\partial p}{\partial z} \rangle}{\langle \theta^2 \rangle} \approx - \begin{cases} \frac{1}{3}\left(1 + \frac{2}{5}\alpha\right) & \text{in the thermal isotropy limit } (\alpha \ll 1) \\ 1 - \frac{\pi}{2\sqrt{\alpha}} & \text{in the infinite thermal unisotropy limit } (\alpha \gg 1). \end{cases} \tag{A5}$$

Accordingly, the coefficient $C_\theta$ = {1 + [r.h.s. of Eq. (A5)]} turns into 2/3 in the thermal isotropy (corresponding to the neutral stratification) and diminishes in an imaginary case of the infinite thermal anisotropy. Our empirical estimate, Eq. (63), of $C_\theta$ = 0.3 is a reasonable compromise between these two extremes.

# Figure captions

**Figure 1.** Ri-dependences of (a) turbulent Prandtl number, $\Pr_T = K_M / K_H$, and (b) flux Richardson number, $\mathrm{Ri}_f = -\beta F_z (\tau S)^{-1}$, after meteorological observations: slanting black triangles (Kondo et al., 1978), snow-flakes (Bertin et al., 1997); lab experiments: black circles (Strang and Fernando, 2001), slanting crosses (Rehmann and Koseff, 2004), diamonds (Ohya, 2001); LES: triangles (Zilitinkevich et al., 2007); DNS: five-pointed stars (Stretch et al., 2001). Solid lines show our model for homogeneous turbulence; dashed line, analytical approximations after Eq. (64).

**Figure 2.** Same as in Figure 1 but for the squared dimensionless turbulent fluxes of (a) momentum, $\hat{\tau}^2 = (\tau/E_K)^2$, and (b) potential temperature, $\hat{F}_z^2 = F_z^2/(E_K E_\theta)$, after lab experiments: diamonds (Ohya, 2001) and LES: triangles (Zilitinkevich et al. (2007); and meteorological observations: squares [CME = Carbon in the Mountains Experiment, Mahrt and Vickers (2005)], circles [SHEBA = Surface Heat Budget of the Arctic Ocean, Uttal et al. (2002)] and overturned triangles [CASES-99 = Cooperative Atmosphere-Surface Exchange Study, Poulos et al. (2002), Banta et al. (2002)].

**Figure 3.** Same as in Figure 2 but for re-normalised turbulent fluxes: (a) $\hat{\tau}^2/(\hat{\tau}^2)^{(0)}$ and (b) $\hat{F}_z^2/(\hat{F}_z^2)^{(0)}$, where the superscript (0) indicates mean values at Ri=0 [hence $\hat{\tau}^2/(\hat{\tau}^2)^{(0)}$ and $\hat{F}_z^2/(\hat{F}_z^2)^{(0)}$ turn into unity at Ri = 0].

**Figure 4.** Same as in Figure 2 but for the vertical anisotropy of turbulence, $A_z = E_z/E_K$, on addition of DNS data of Stretch et al. (2001) shown by five-pointed stars.

**Figure 5.** Gradient Richardson number, $\mathrm{Ri} = \beta(\partial\Theta/\partial z)(\partial U/\partial z)^{-2}$, versus dimensionless height *z/L* in the nocturnal atmospheric boundary layer (ABL). Dark- and light-grey points show LES data within and above the ABL, respectively; heavy black points with error bars are bin-averaged values of Ri [from Figure 3 of Zilitinkevich and Esau (2007)]. Solid line is calculated after Eqs. (41), (55) and (65) with *n* = 4/3.

**Figure 6.** Dimensionless vertical gradients of (a) mean velocity, $\Phi_M = \dfrac{kz}{\tau^{1/2}}\dfrac{\partial U}{\partial z}$, and (b) potential temperature, $\Phi_H = \dfrac{k_T z \tau^{1/2}}{-F_z}\dfrac{\partial\Theta}{\partial z}$, versus *z/L*, after our local closure model [solid lines plotted after Eq. (5.13a,b)] compared to the same LES data as in Figure 5 [Figures 1 and 2 of Zilitinkevich and Esau (2007)].

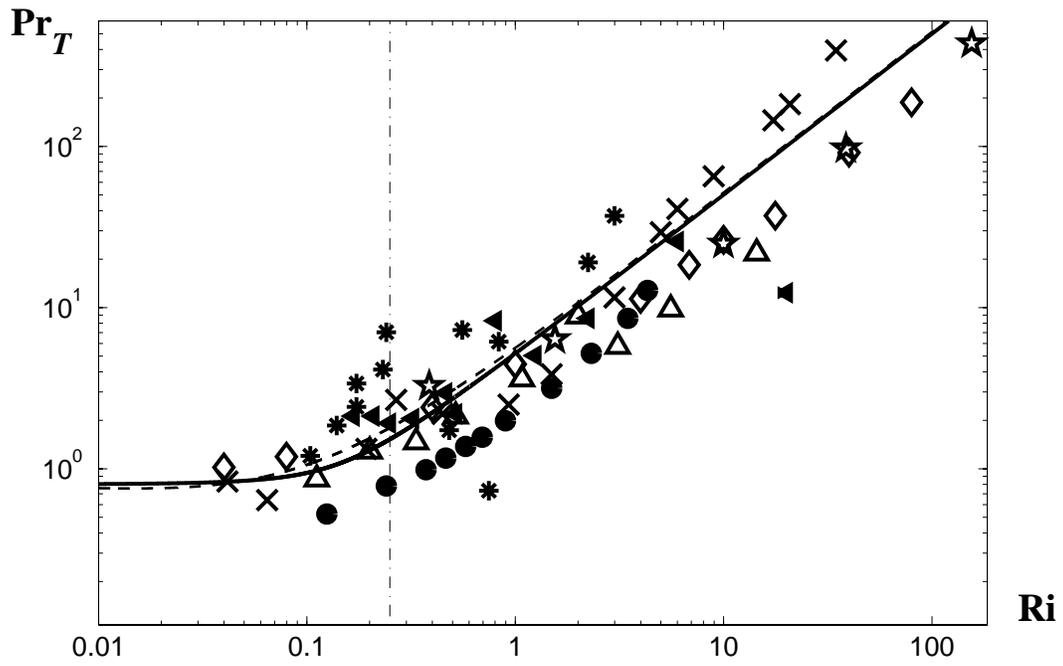

Figure 1(a)

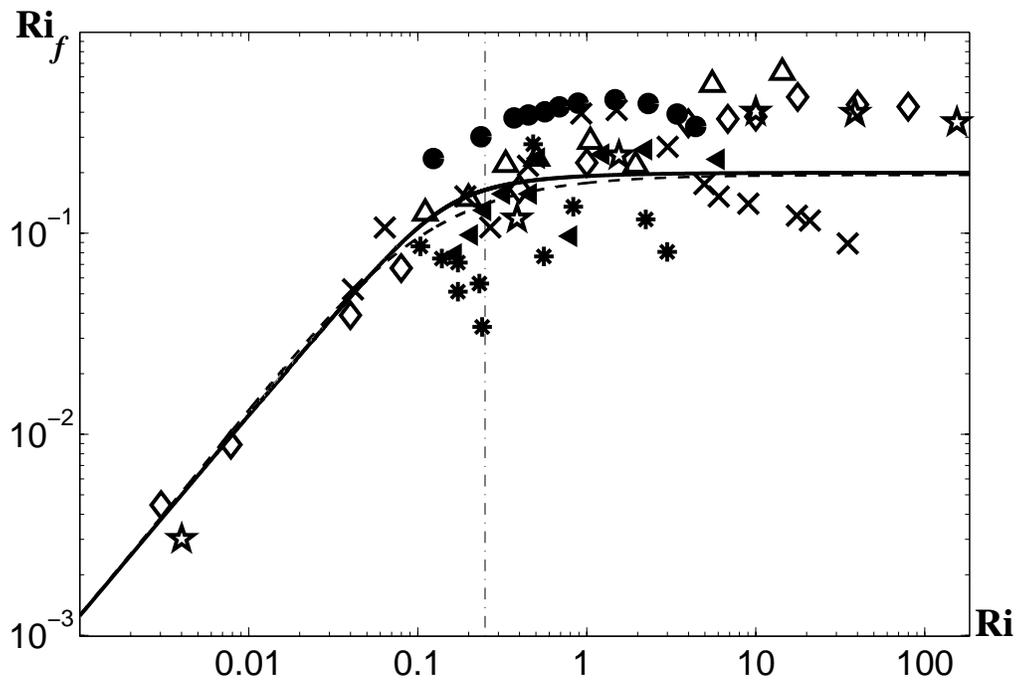

Figure 1 (b)

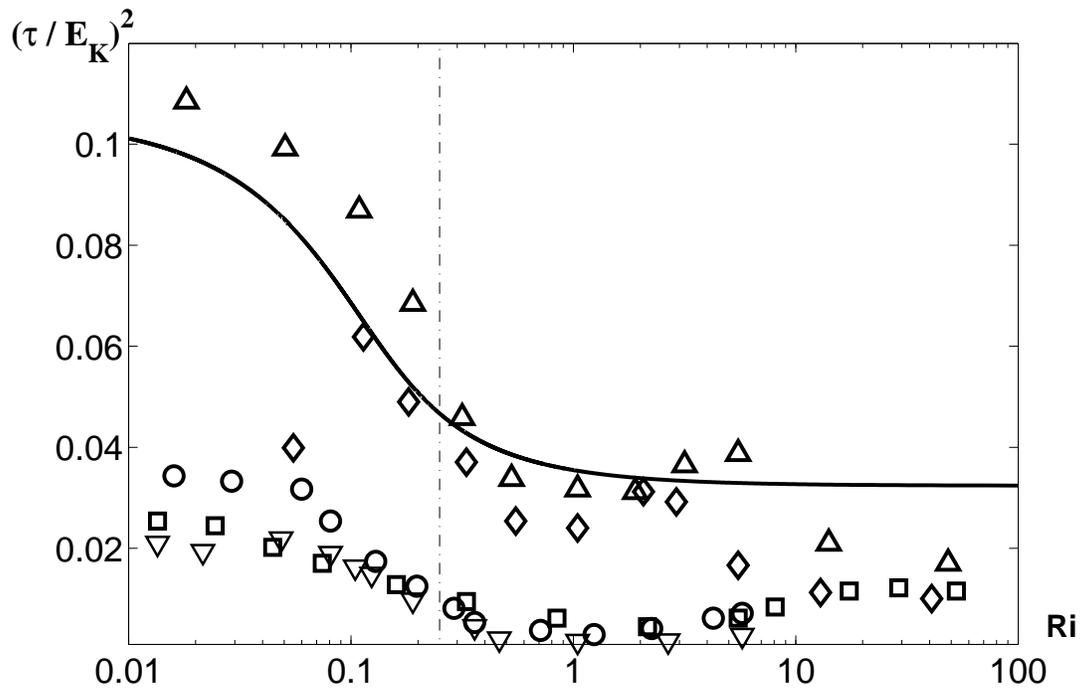

Figure 2 (a)

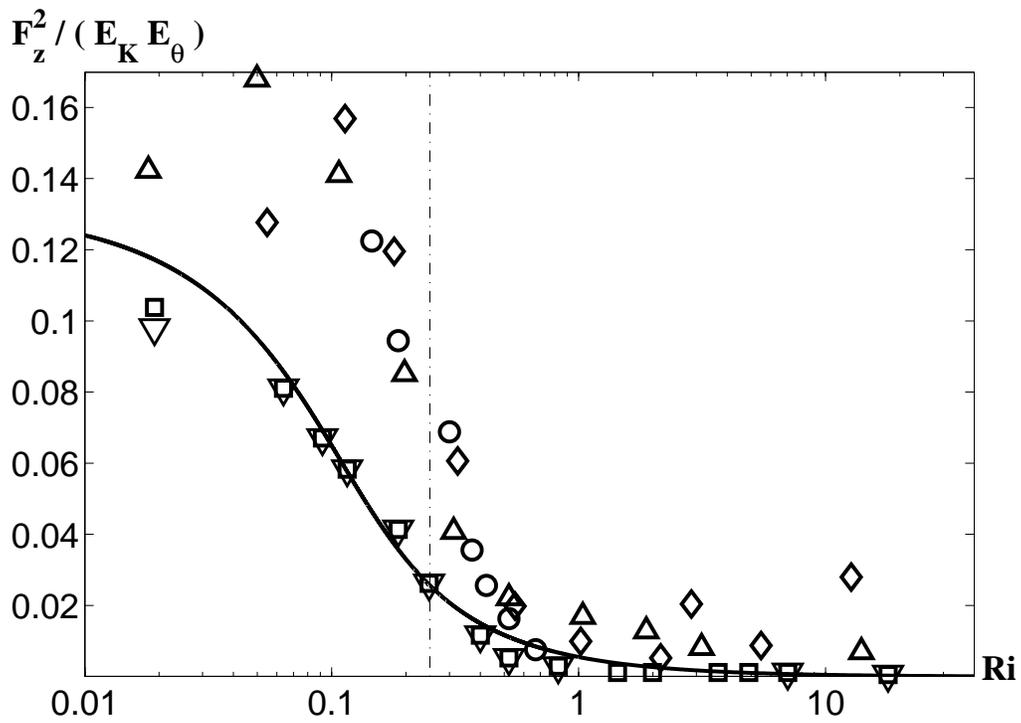

Figure 2 (b)

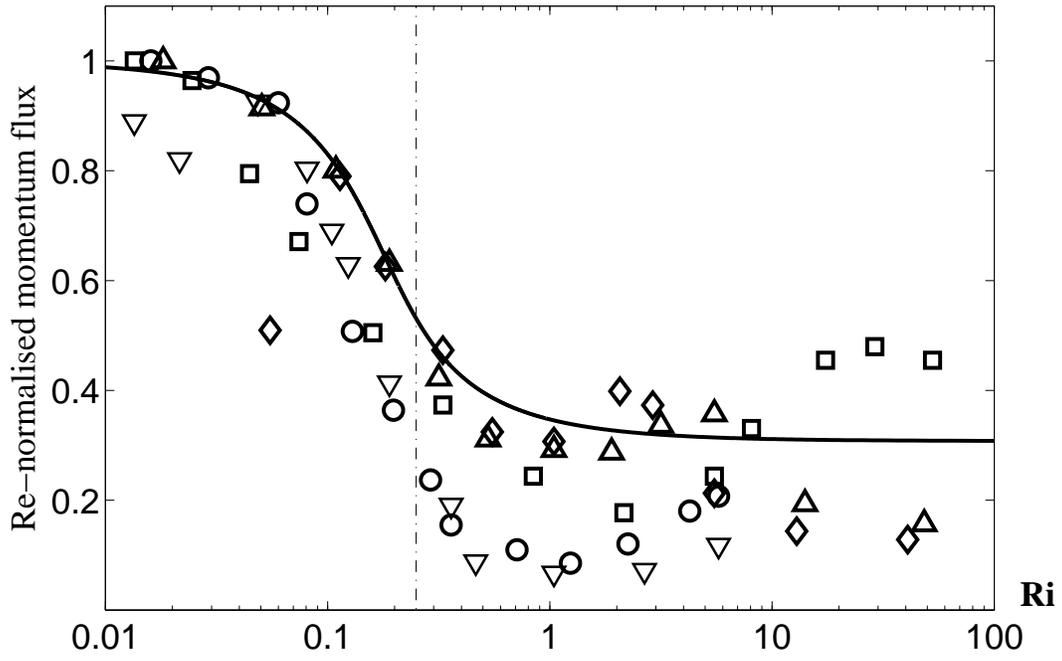

Figure 3 (a)

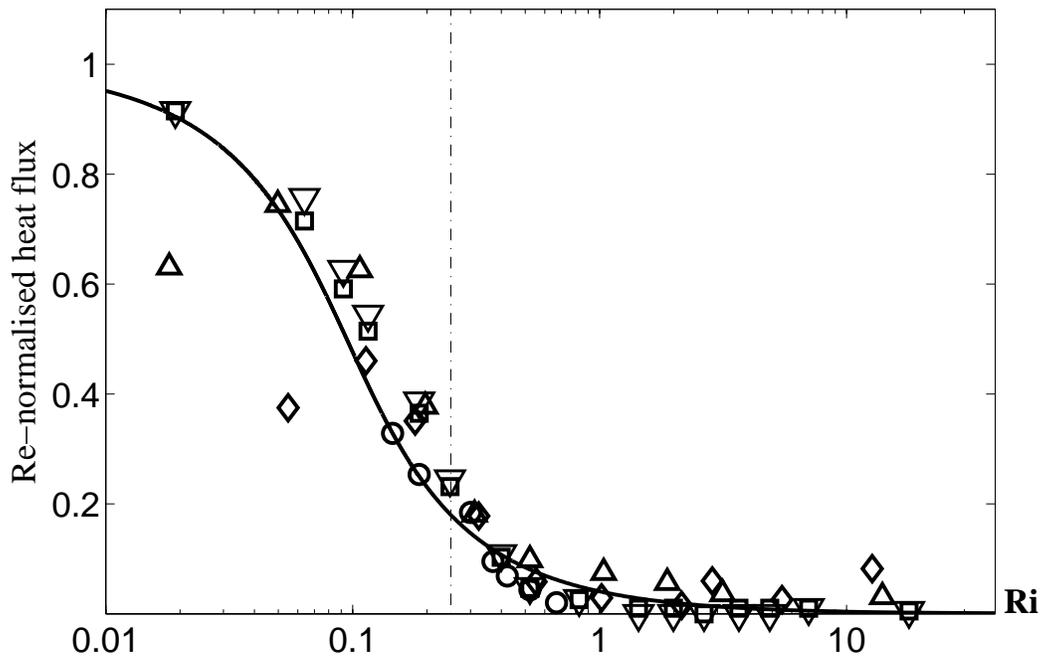

Figure 3 (b)

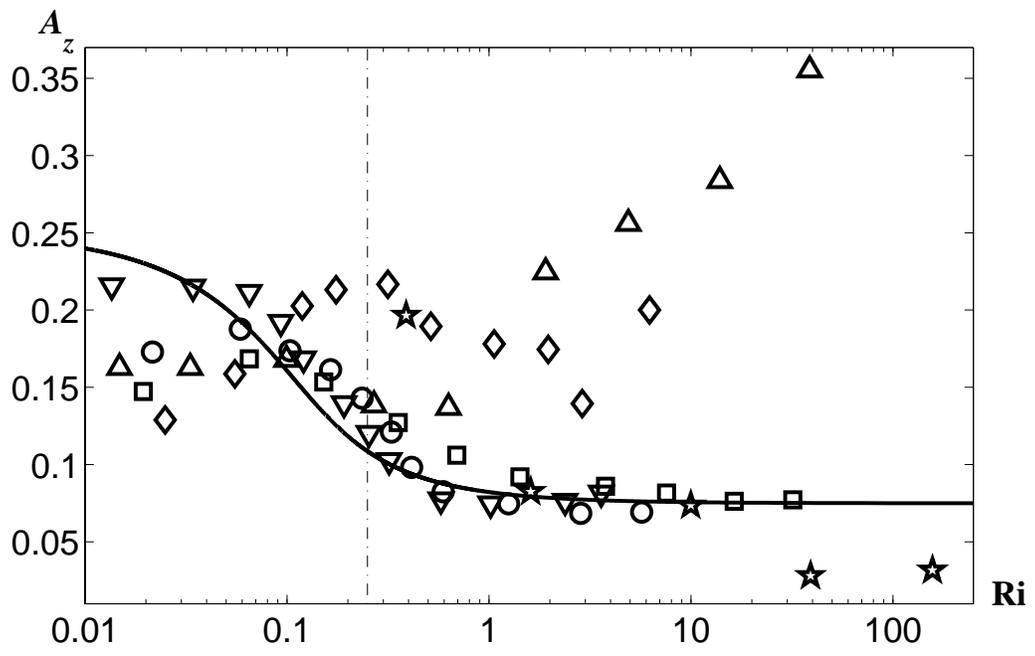

Figure 4

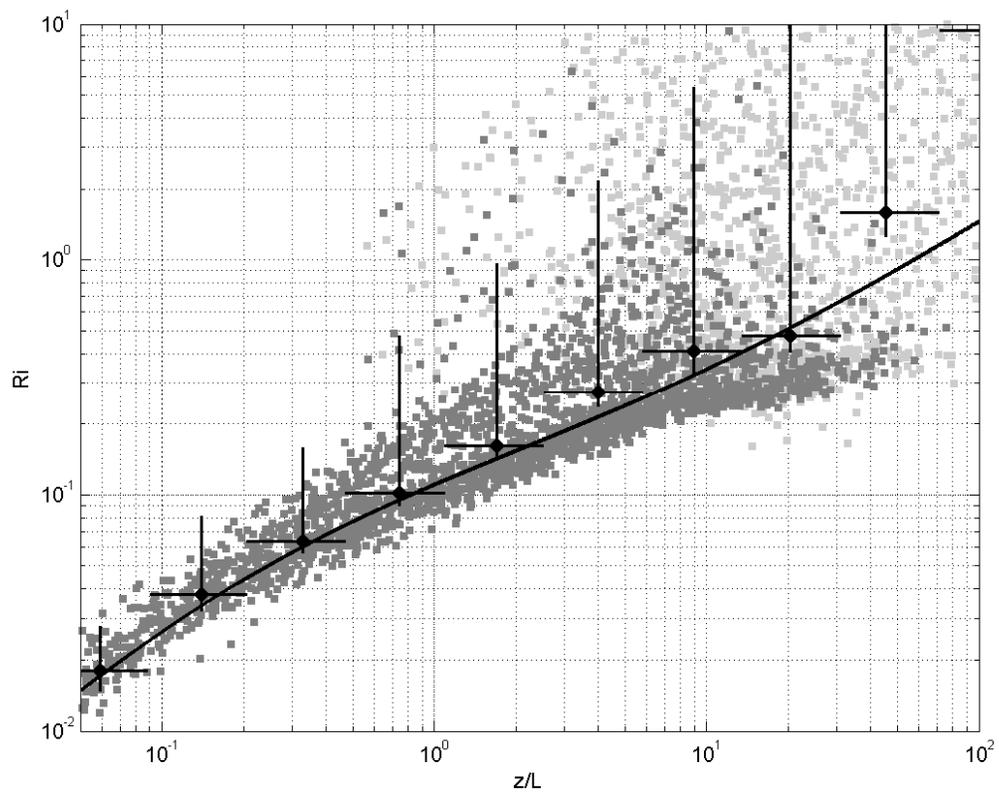

Figure 5

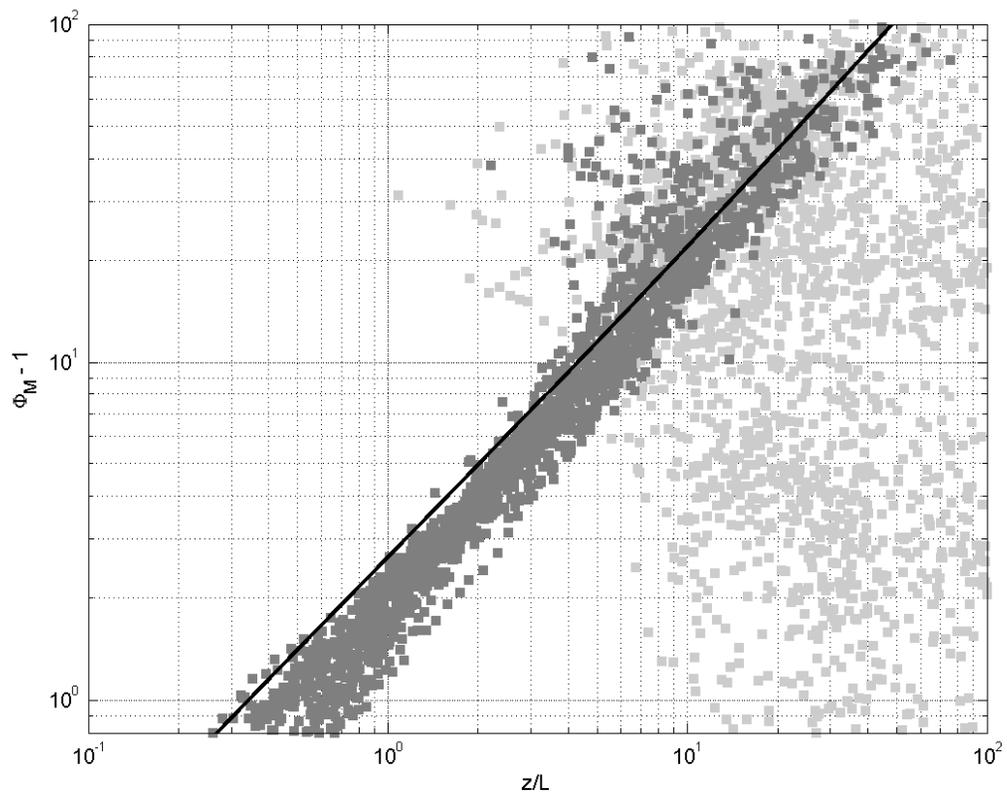

Figure 6a

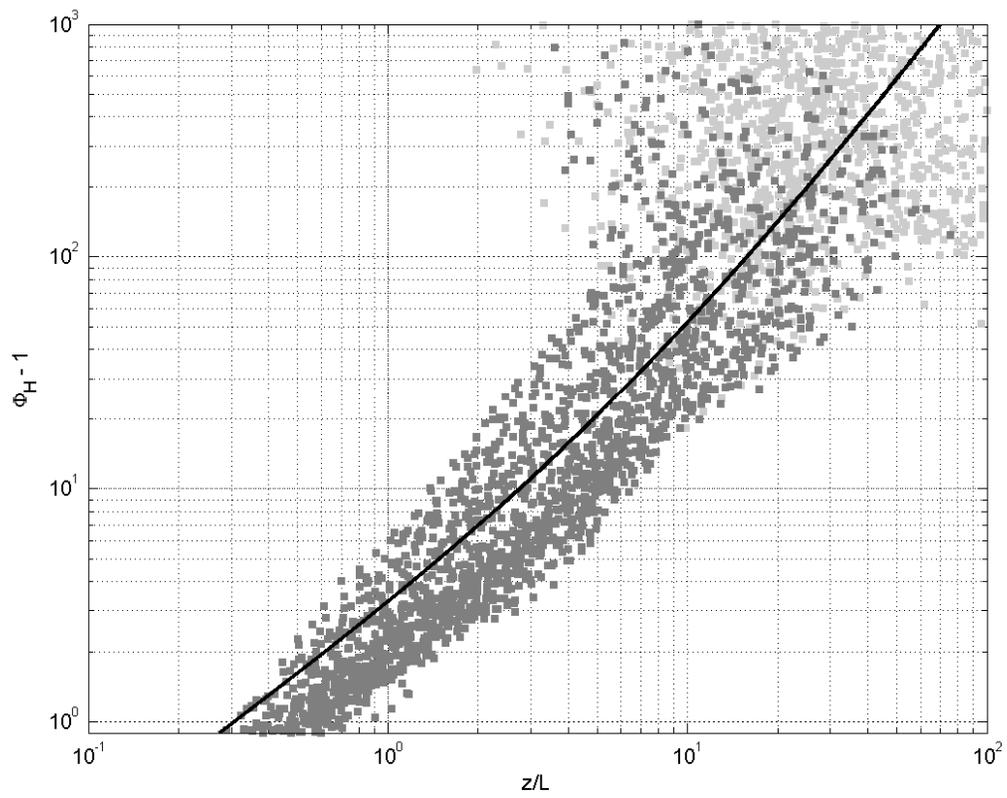

Figure 6b